\documentclass[a4paper,oneside]{article}

\usepackage[dvips]{graphicx}

\usepackage{amsmath}
\usepackage{amsfonts}
\usepackage{amssymb}
\usepackage{amsthm}
\usepackage{amscd}
\usepackage[mathscr]{eucal}

\usepackage{hyperref}

\newcommand{\myURL}[1]{\url{#1}}

\newcommand{\myRef}[1]{
  \S\ref{#1}
}

\newcommand{\eg}{\textit{e.g.}\,}
\newcommand{\ie}{\textit{i.e.}\,}

\newcommand{\Java}{\emph{Java}\,}

\newcommand{\mySrc}[1]{\textsf{#1}}

\newlength{\facewd} \newlength{\faceht}

\begin{document}

\title{ODP channel objects that provide services transparently for
  distributing processing systems}

\author{Walter D Eaves}


\date{\today}

\maketitle

\begin{abstract}
  This paper describes an architecture for a distributing processing system
  that would allow remote procedure calls to invoke other services as
  messages are passed between clients and servers. It proposes that an
  additional class of data processing objects be located in the software
  communications channel. The objects in this channel would then be used to
  enforce protocols on client--server applications without any additional
  effort by the application programmers.  For example, services such as
  key--management, time--stamping, sequencing and encryption can be
  implemented at different levels of the software communications stack to
  provide a complete authentication service. A distributing processing
  environment could be used to control broadband network data delivery.
  Architectures and invocation semantics are discussed, Example classes and
  interfaces for channel objects are given in the \Java programming
  language.
\end{abstract}

\section{Distributed Processing Platforms}

\subsection{RPCs and Connections}

A distributed processing platform provides a method of making a Remote
Procedure Call, an RPC, almost transparent to the application
developer---see, for example, \textit{Tivoli}\cite{orb:tivoli} and
\textit{Orbix} \cite{orb:orbix}---the relatively mature industry standard
for both of which is the Common Object Request Broker Architecture,
\textit{CORBA}, from the Object Management Group\cite{web:corba}, OMG; this
defines an Object Request Broker, ORB, which is an infrastructure for RPCs.

Recently \textit{Sun Microsystems} \cite{web:sun} have enhanced the Remote
Method Invocation package, \mySrc{java.rmi}, for \Java\cite{web:java} as
part of the \Java Development Kit, JDK, 1.2. It now allows connections to
be created between a client and a server which can have a different data
transfer representation \cite{java:rmi:sockets}. As pointed out in the
documentation for this feature, this is particularly suitable for
implementing the Secure Socket Layer, SSL, \cite{draft:ssl} and could also
be used to implement the proposed successor to SSL Transport Level
Security, \cite{draft:tls}.

\paragraph{Open Distributed Processing Architecture}

A suitable architecture to exploit this new functionality in
\mySrc{java.rmi} and in other distributed processing platforms has been
proposed in the Open Distributed Processing standards, ISO--ODP,
\cite{ISO:ODP}.

\subparagraph{Bindings and Channels}

\begin{figure}[htbp]
  \begin{center}
  \includegraphics[angle=-90, keepaspectratio=1, totalheight=6in]{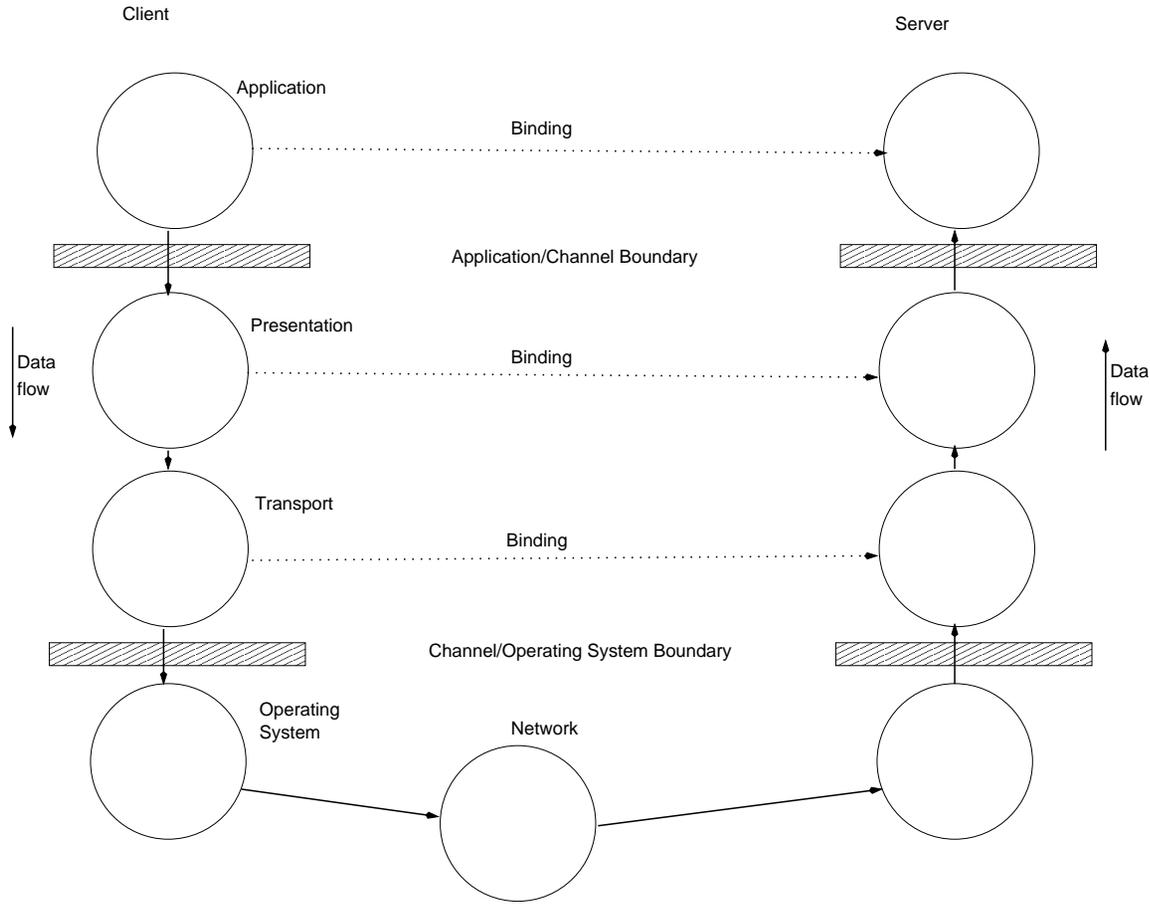}
  \caption{Bindings across two channels}
  \label{fig:odp-channel}
  \end{center}
\end{figure}

The prescriptive model, \cite{ODP:presc}, generalizes the concept of
connection between client and server to be a logical binding which is
realized by both client and server using a channel, see also
\cite{sft:ansaware}. Figure \ref{fig:odp-channel} illustrates the concepts
of bindings and channels.  Application objects have bound with one another
probably using a naming service; they have negotiated a channel
configuration that requires two objects: a data presentation conversion
object and a data transport object.  The channel configuration is fixed by
bindings. The client sends data; the server receives. The data is passed
from the application object to the channel which carries out the conversion
and the transport to the server.

\begin{description}
\item[Binding] A binding is a contract between the client and the server
  stating the parameters by which they will communicate. A binding between
  two transport objects would usually specify:
  \begin{itemize}
  \item Transport layer protocol to be used, \eg UDP/IP or TCP/IP or
    Pipe/Unix.
  \item Transport layer parameters
  \item Network addresses for chosen transport layer protocol
  \end{itemize}
  
  The presentation layer object would convert from the the local data
  representation to a network representation. The only specification needed 
  here is the source and target representations. In this case, the binding
  is implicit in the implementation of the objects, they need not be
  initialized with parameters.
  
  More sophisticated systems will have higher demands and will insist that
  other services be used as well, for example:
  \begin{itemize}
  \item Data security
  \item Transaction management
  \item Call billing
  \item Data compression
  \item Relocation manager
  \end{itemize}
  These would all require that configuration parameters be specified and
  may also demand that they are operate together.
\end{description}

There is no need for channels to be symmetric, the server could implement a 
call--logging object in its channel without having an object of the same
type in the client's channel. 

Bindings do need to be current. A transport object might close a
connection, in which case, it would no longer be current, but if it were to
leave enough information to allow a re--connection, then it is, in effect,
still current.

\subparagraph{Tri--partite Bindings}

\begin{figure}[htbp]
  \begin{center}
  \includegraphics[keepaspectratio=1, totalheight=4in]{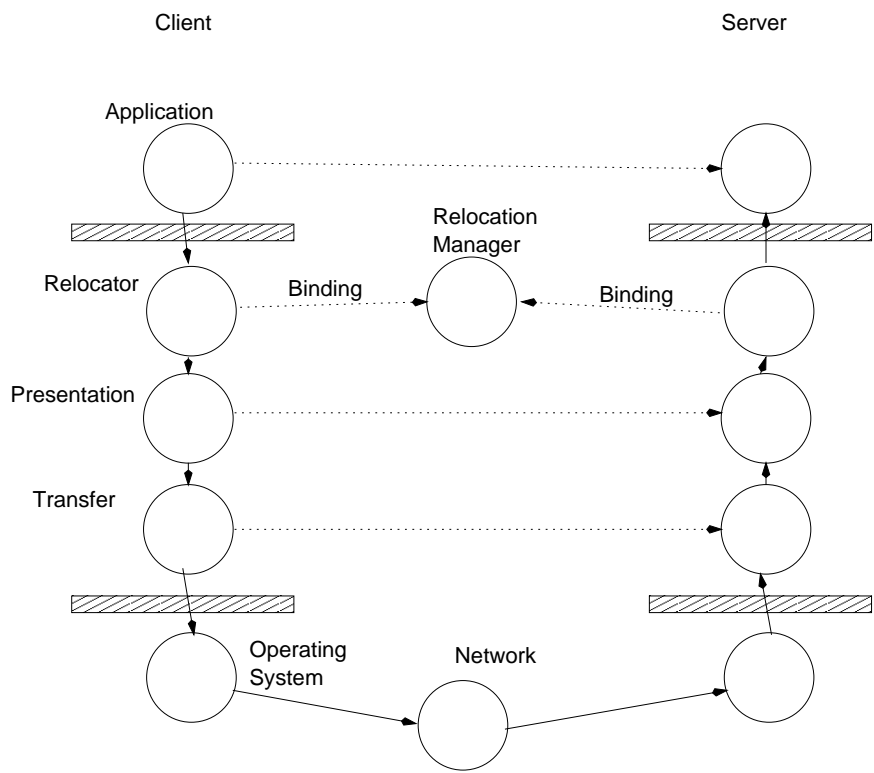}
  \caption{Relocation: a tri--partite binding}
  \label{fig:odp-3binding}
  \end{center}
\end{figure}

Bindings need not be bi--partite. An application service that would require
a tri--partite binding is relocation management, illustrated in figure
\ref{fig:odp-3binding}. The idea of which is that should the server choose
to relocate, it would notify a relocation manager of its new addresses and
move there. When the client calls the server at its old address and fails
to reach it, the relocator object in the client's channel would call the
third party and ask for the new address of the server, establish a new set
of bindings with it, \ie construct a new channel and destroy the old, and
send the message again. This would all be transparent to the application
object.

The relocation object in the channel would need to call the relocation
manager for the new addresses and would thus become an application object;
it would need to establish its own channel with the relocation manager.

\subsection{Inflexible System Designs}
\label{sec:inflexible}

Although the ISO--ODP defines a flexible architecture, most implementors of
distributed processing platforms provide application programmers with
inflexible systems: the channel can only contain a presentation and a
transport object.

\paragraph{Presentation: Stubs and Skeletons}

Both \textit{CORBA} ORBs and the RMI defined in \Java make use of ``stubs''
and ``skeletons'' and a stub--compiler. The term infrastructure will be
used for the engineering that realizes an ORB or RMI.

\begin{description}
  
\item[Stubs] These are used by a client when invoking a method on a remote
  server. The client does not have a local implementation for the service
  the remote server has, but it will have a definition of its interface
  which can be used as if it were the implementation of the service. It can
  present the interface definition to a stub--compiler which will generate
  code to invoke each method on the remote server. The stub acts as a proxy
  for the remote server in the client's address space. The stub for each
  method needs to do the following things:
  \begin{enumerate}
  \item Construct a request

    A request object is a container entity that carries the invocation to
    the server; it contains:
    \begin{itemize}
    \item The address of the remote server
    \item The name of the method being invoked, possibly with
      version control information for the interface.
    \item The parameters for the method
    \item The return address
    \end{itemize}
    
    The stubs can also contain the code to \emph{marshall}\footnote{The
      verb is \textit{to marshal}, but this is so often mis-spelt that to
      marshall has become acceptable.} the parameters into a universal
    transfer presentation. In this form, the stubs also comprise the
    presentation object.
    
  \item Invoke the request
    
    The request, now just a sequence of bytes with an associated address
    for the server and a return address for the sender, is passed to the
    infrastructrure which sends it to the network socket for the server.

  \item Get the reply

    The server will return a reply in another container type.
    
  \item Re--construct the reply
    
    The reply is then re--constructed or, rather, its contents are
    \emph{un--marshalled} and returned to the client. Again, this may be
    part of the code in the stubs.
     
  \end{enumerate}

\item[Infrastructure] It might be best now to explain how the
  infrastructure manages to provide the RPC service.
  
  \begin{enumerate}
  \item At the server on creation
    
    An application programmer defines an interface and produces an
    implementation for it. A program that effectively acts as a loader
    issues instructions to the infrastructure to create a socket to receive
    requests for that server and will associate the remote server
    implementation with that socket.

  \item At the naming service
    
    The application programmer will ensure that the address of his
    newly--created remote server is put into a well--known naming
    service. The client will then collect the address from the naming
    service.
    
  \item At the client on invocation
    
    The address collected at the naming service will contain enough
    information to allow the client's infrastructure to send the request
    container as a stream of bytes to the socket that the server's
    infrastructure has associated with the remote server's implementation.

  \item At the server on invocation
    
    The network socket will be activated by the client's infrastructure (a
    connect and a data send) and the server's infrastructure will collect
    the data at the socket and, because it has recorded the object
    responsible for that socket, it can activate the server skeleton.
    
  \end{enumerate}
  
\item[Skeletons] \label{sec:inflexible:1} These are invoked by the server's
  infrastructure when the network socket for a server is activated and the
  data comprising a call has been collected. It invokes the implementation
  of the method the client wants to use at the server. A skeleton can
  consist of one method that unmarshalls enough of the request container to
  be able to look up the method to invoke---this is known as
  \emph{dispatching}.
  
  This partial unmarshalling has to be done in this way, because each
  method will unmarshall the remainder of the request container differently
  to obtain the parameters to pass to the server's method implementation.

  After the invocation is made the results will be marshalled into a reply
  container which is then sent back to the client.
  
\end{description}

The stubs and the skeleton effectively form the presentation channel
object. Stubs have to be produced by a stub--compiler, but skeletons can be
made generic, if the underlying infrastructure supports a reflective
invocation mechanism, see \mySrc{java.lang.reflect} or the \textit{CORBA}
Dynamic Invocation Interface.

Some stubs generated by the \Java \mySrc{rmic}, Remote Method Interface
Compiler, are given in appendix \ref{cha:stubs}. These demonstrate the use
of reflective language features.

\paragraph{Transport}

When the request containers are passed to the infrastructure the transport
object is eventually invoked. Most ORBs only provide one transport
mechanism which sends data through TCP/IP sockets to its destination.
Although some systems do allow different protocols: UDP/IP or Pipe/Unix.

\paragraph{Limitations}

\begin{figure}[htbp]
  \begin{center}
  \includegraphics{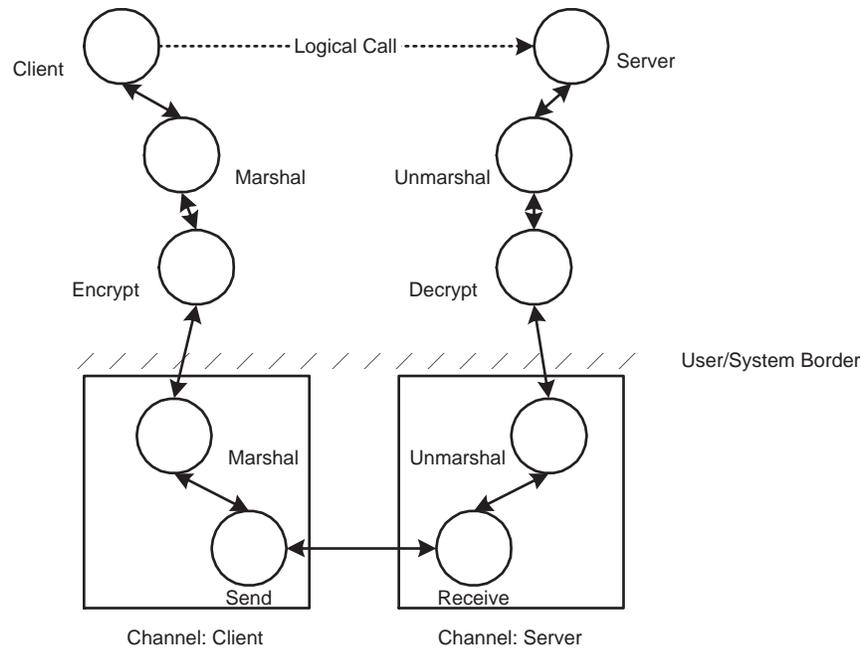}
  \caption{Encrypting and Decrypting by the application programmer}
  \label{fig:ODP-channel-0}
  \end{center}
\end{figure}

The problem with this is that if one wants to implement any useful
application services---encryption, billing and so forth---the
infrastructure does not help. For example, to encrypt and decrypt data sent
as part of a remote procedure call, one would have to implement one's own
stubs and skeletons, see figure \ref{fig:ODP-channel-0}.

The application programmer has to construct a call, marshall the data,
encrypt it, send it using a generic method, which will marshall it again.
At the server, the data would be delivered to the generic method, and the
application programmer would then have to decrypt it, unmarshall the
decrypted data, reconstruct the call and dispatch it. After dispatching,
collect the results, marshall, encrypt and return the reply.

\subsection{More flexible: \Java RMI Custom Socket Factories}

A more flexible implementation has been provided by \textit{Sun} in
\Java. It allows a different type of socket to be used as the transport
object.

\paragraph{Method}

The custom socket factories have to implemented in the following way:

\begin{enumerate}
\item Derive and implement classes for the new socket type's datastream
  from \mySrc{java.io.FilterOutputStream} and
  \mySrc{java.io.FilterInputStream}, call them \mySrc{MyOutputStream} and
  \mySrc{MyInputStream}.
\item Derive and implement classes for the new socket types
  \mySrc{java.net.Socket} and \mySrc{java.net.ServerSocket} that use the
  new streams \mySrc{MyOutputStream} and \mySrc{MyInputStream}.
\end{enumerate}

Then create socket factory implementations that can be used by RMI.

\begin{enumerate}
\item A client-side socket factory that implements
  \mySrc{RMIClientSocketFactory} and implement the
  \mySrc{RMIClientSocketFactory.createSocket()} method.
  
\item A server-side socket factory that implements
  \mySrc{RMIServerSocketFactory} and implement the
  \mySrc{RMIServerSocketFactory.createServerSocket()} method.
\end{enumerate}

Then one has to ensure that the constructor for the remote server is told
to use the new socket factories. The infrastructure creates the new type of
socket when demanded and invokes the create socket methods.

The \mySrc{RMISecurityManager} at the client will then determine that a
particular type of socket has to be used and will load the custom socket
factory implementations.

\paragraph{Possibilities: Implementing SSL}

Using custom socket factories, it is possible to implement a Secure Socket
Layer. The \mySrc{RMIClientSocketFactory.createSocket()} would be used to
perform the key exchange with the server and the custom input and output
streams would apply the session key to encrypt on send and decrypt on
receive.

\paragraph{Limitations}

Unfortunately, using custom socket factories only increases the variety of
the transport objects that can be employed, it does not allow different
kinds of objects to be placed in the channel.

\section{Channel Objects}

What is needed is a means of placing objects in the channel before and
after the presentation object. These objects should have a simpler
instantiation and invocation procedure than using the custom socket method
in \Java.

\subsection{Some Requirements}

\begin{enumerate}
\item Different interests \- System Configurable
  
  The objects placed in the channel between client and server are the
  result of a negotiated agreement between the client, its server and their
  respective environments. It is well--known that security requirements for
  messages depend on the workstation that the client is using
  \cite{sec:sesame}, which may be connected to a secure local area network
  on which both the the client and server reside and so, for example, no
  security measures need be taken; or, the client could be accessing the
  server remotely from the Internet through a modem in which the server
  might require that the client use encryption.
  
  It would be desirable if the client and the server could both specify
  their requirements and some negotiation take place that could create a
  mutually acceptable protocol stack in the software communications
  channel.
  
\item Different Methods for Different Actions
  
  A request and reply actually require that four different channels be
  traversed by messages, see figure \ref{fig:odp-4channels}:
  \begin{enumerate}
  \item Request \- this channel is created by the infrastructure for the
    client and sends a message over the network.
  \item Indication \- this channel is used to receive from the network and
    is created by the infrastructure for the server.
  \item Response \- created by the infrastructure for the server to return
    the results of the client's request message.
  \item Confirmation \- receives from the network and is created by the
    infrastructure for the client.
  \end{enumerate}
  
  Should an error occur at the server it is returned via the response and
  confirmation messages.
  
  The stubs are responsible for managing the thread of execution of the
  application object: once a message is put onto the request channel, the
  application object's thread can be suspended whilst it waits for the
  reply to arrive on the confirmation confirmation channel.
  
  If the message is a cast of some kind (broadcast, multi--cast or
  one--cast) there will be no response or confirmation.

  \begin{figure}[htbp]
    \begin{center}
      \includegraphics[angle=-90, keepaspectratio=1, totalheight=6 in]{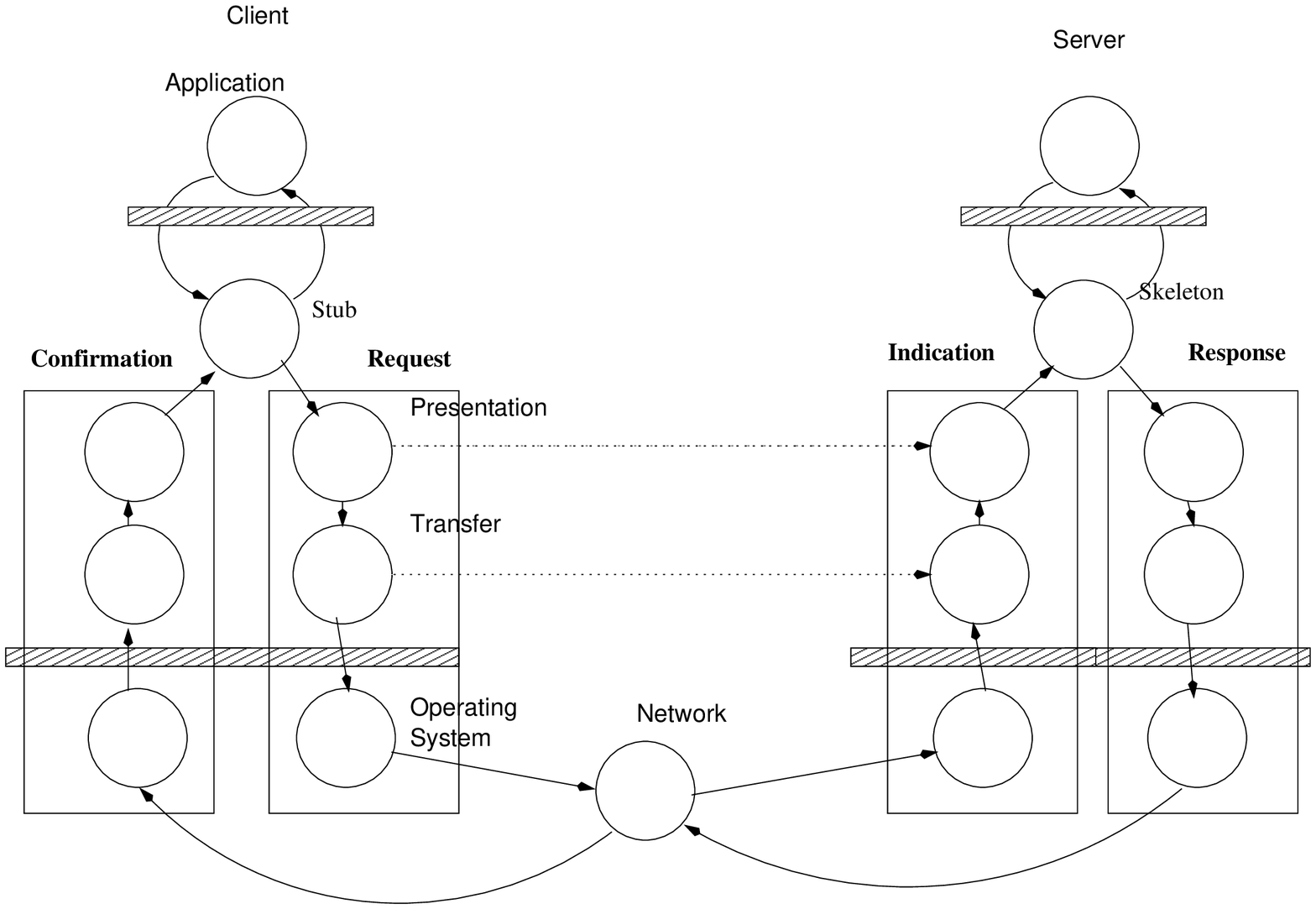}
      \caption{Four channels are used in a request and response}
      \label{fig:odp-4channels}
    \end{center}
  \end{figure}

\item Different Implementations
  
  Most channel objects are derived from the same source and will have the
  same implementation. \Java allows classes implementing different objects
  to be loaded over the network, so it would be possible for the client and
  server to agree upon and load the same class implementation, which they
  would be able to do with the custom socket factories method. This may not
  always be the case, some channel objects may be optimized to make use of
  a different operating system, but provide the same functionality.  Audio
  and video data streaming are good examples of this need: some
  micro--processors now have support for stream data.
  
\item Transparency

  \begin{enumerate}
  \item Management
    
    It would be desirable if the additional services provided by the
    channel objects did not need to be initialized or managed by either the
    client or the server application programs but they were activated by
    their respective infrastructures.
    
  \item Exception handling
    
    One of the difficulties of developing applications in enterprise
    environments is that as messaging becomes more
    sophisticated---supporting for example, confidentiality, authorization,
    call--billing---the number of possible errors increases because each of
    these sub--systems introduces new ones. It must be possible for channel
    objects to clear down their own errors, so that errors returned to
    application objects only involve the application.
    
  \item Using Possession rather than Inheritance for Coupling
    
    It would also be desirable if the classes for the channel objects did
    not extend the existing class hierachies of the message transmission
    sub--system, in the way that custom socket and socket factories
    do. Extending class hierachies is not as flexible as specifying an
    order of invocation.
    
  \item Interface Definitions Suitable for Reflection.
    
    It would also be desirable if the infrastructures could load channel
    objects' classes remotely and have a simple enough invocation syntax so
    that reflection mechanisms could be used to invoke the channel objects
    \textit{without requiring a stub compiler}. \Java already does this
    with its version 1.2 stubs, using package \mySrc{java.lang.reflect},
    and \textit{CORBA} has a Dynamic Invocation Interface which can achieve
    the same goal.
    
  \end{enumerate}
  
\item Efficiency
  
  It would also be desirable if the channel objects did not create a large
  stack of calls; primarily because some target environments for remote
  procedure call platforms will be embedded systems on SmartCards
  \cite{smartcard:tech}.
  
\end{enumerate}

\subsection{Some Nomenclature}

With the aid of figure \ref{fig:odp-4channels}, it is possible to be more
precise with the terms used:

\begin{itemize}
\item Initiator: starts a four--phase call sequence; may also be called the
  requestor.
\item Acceptor: accepts the call made upon it; may also be called the
  responder, because it generates the response.
\end{itemize}

Both an initiator and an acceptor will send and receive as part of
four--phase call sequence. Ordinarily, the client will be the initiator of
all calls, but some remote procedure call systems permit call--backs, in
which case the server is the initiator and the client the acceptor.

\section{Architectures for Channel Objects}

There are basically two types of architecture that could support channel
objects.

\subsection{Stream--oriented Architecture}
  
The custom socket factory architecture is stream--oriented. It acts upon
the data being sent between client and server as a stream applying a data
transformation to the data when it is sent and undoing this transformation
at the other end.
  
The data is treated as opaque and can be delivered in packets as small as
one byte. The implementor of the stream handler does not know whether the
data has just started or is about to finish.
  
One of the attractions of this approach is that many of the operations that
data networks perform on data can be implemented in software: segmenting,
and its converse re--assembling, can be implemented easily and this would
allow remote procedure call systems to make use of packet--oriented
transmission, such as UDP/IP. Segmenting and re--assembly could be
implemented as channel objects and data would be segmented and then sent as
packets on a UDP socket for re--assembly by another channel object at the
server.
  
Streams can be chained---the output of one providing the input for the
next. Most operating systems support \textit{pipes} to do this: \Java
supports the \mySrc{java.io.PipedInputStream}, and output, classes.
  
Multi--casting could be easily achieved with pipes: simply have a pipe that
sends on a socket and echoes its input; then connect a series of these
together.
  
A stream--oriented architecture is better for real--time data delivery.
All of the processing in the stream--handler is applied to the data---it is
not expected to communicate with relocation or transaction managers and
incur indeterminate time penalties. Consequently, the emphasis in the
design of stream--handlers should be to ensure they introduce a constant
latency in transmission and reception.
  
\subsection{Call--oriented Architecture}
  
This architecture implicitly appreciates that the data being delivered is a
call.
  
Call--handlers usually add parameters to a remote procedure call. They do
not form part of the message sent by the application object, but set its
context. For example:
\begin{itemize}
\item Timestamps \- logging when messages are sent and received by adding
  an ``time--sent'' parameter to the remote procedure call.
\item Accounting \- adding an account identifier to a remote procedure call
  so that the server can log charges for calls to a particular account.
\item Transactions \- a transaction might consist of a number of calls to
  different servers, they could all be identified with a transaction
  identifier, to synchronize committing and aborting transactions as a
  whole.
\item Authorization \- a Privilege Certificate could be attached to the
  call so that the server could check what rights and privileges the caller
  is allowed to exercise within the server's work--space.
\end{itemize}
  
A call--oriented architecture \textit{should} be implemented so that it has
access to the parameters passed as part of the request. If this is the
case, then as well as being able to add parameters, it would be possible to
perform data transformations on parameter values that are part of the call.
For example:

\begin{enumerate}
\item Representation conversions

  \begin{enumerate}
  \item Wholly
      
    The presentation objects implemented in remote procedure call systems
    change the data representation of the parameters of a call so that they
    can be transmitted over the network as a sequence of bytes, with the
    receiver being responsible for converting the byte--sequence to its
    local representation. It may be more efficient to convert to the
    receiver's format before the data is sent so that the server can use
    conversion methods available from its native operating system.
      
  \item Partly
      
    It might be the case that a server has a different data context for a
    particular data type: internationalization of text strings and currency
    formats could be converted prior to transmission.
      
  \end{enumerate}
    
\item Pseudo--Objects
    
  Pseudo--objects are usually legacy systems that can be directly
  controlled by the client's remote procedure call infrastructure, but for
  the sake of uniformity, and to simplify re--engineering and relocation of
  services, they are provided with the same interface as remote servers.
  The operating system used by a distributed processing platform is itself
  a legacy system.
    
  An example of a pseudo--object that application programmers use is the
  database driver provided in the \Java DataBase Connectivity package,
  \mySrc{java.sql}. This pseudo--object implementation establishes and
  drives a connection with the database. Other examples of services that
  could be implemented as pseudo--objects are directory and naming services
  that are available through native operating systems.
    
\item Stream
    
  A call--oriented architecture could also be used to transform data in the
  same way that a stream--oriented architecture could. Encryption,
  compression and checksum insertion could all be performed by marshalling
  the parameters using a data representation object to produce a sequence
  of bytes and then applying the stream operation to it. The output would
  be opaque and would replace the parameters.

\end{enumerate}

A call--oriented architecture is better suited to recovering from errors,
since it is possible to determine which channel object is at fault and it
can take measures to recover from the error.

\subsection{Both architectures are needed}

\begin{figure}[htbp]
  \begin{center}
  \includegraphics[angle=-90, keepaspectratio=1, totalheight=6 in]{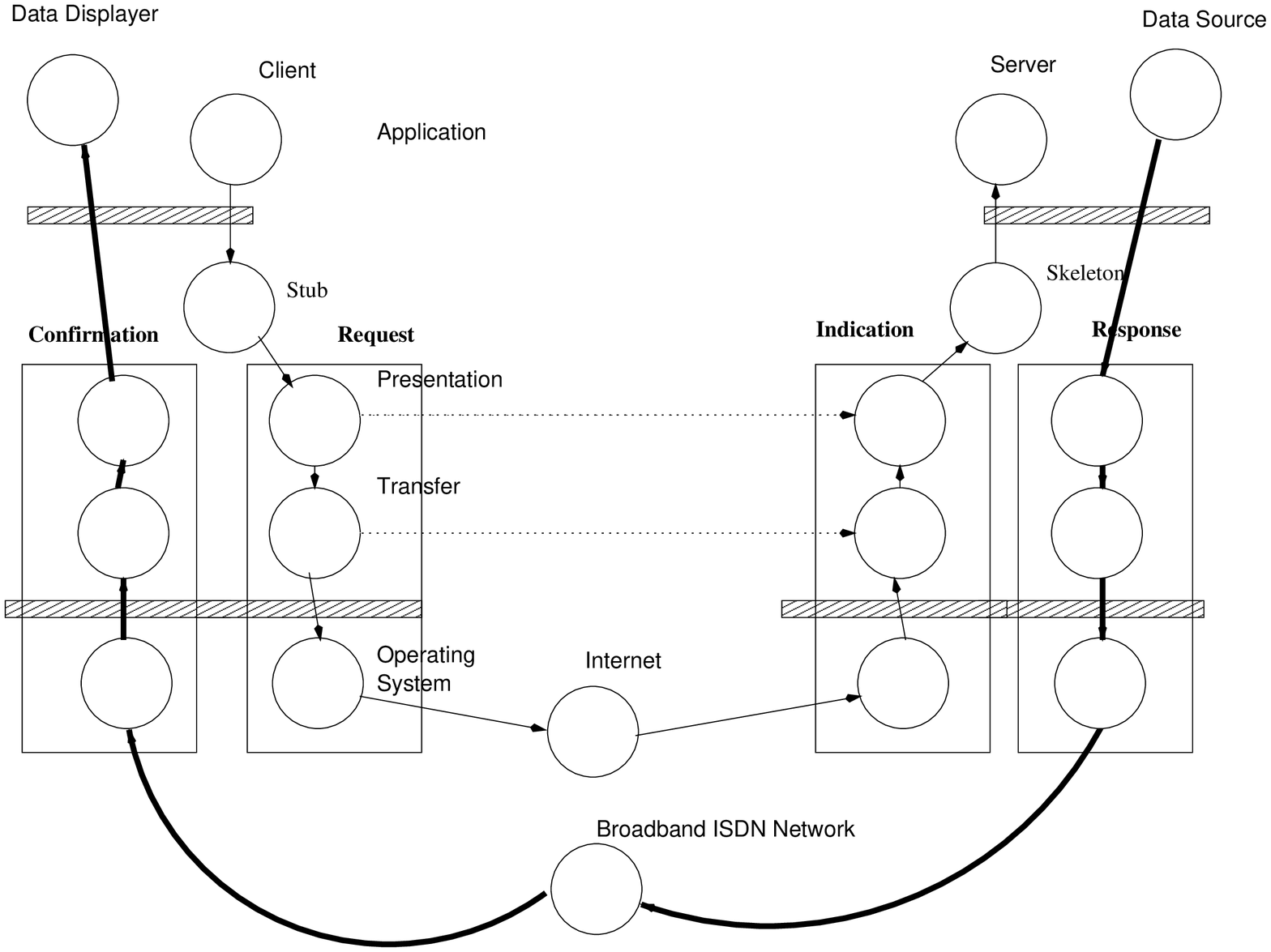}
  \caption{Use of a RPC to connect to a broadband data source}
  \label{fig:odp-2sources}
  \end{center}
\end{figure}

The stream--oriented architecture is ideal for delivering data at high
speed with a determinate latency, the call--oriented architecture is ideal
for communicating control information. This is a similar design problem
that faced the developers of telephone networks and it was resolved, in the
Integrated Services Digital Network, ISDN \cite{IEEE_1990}, by having a
control channel manage the use of two bearer channels. Figure
\ref{fig:odp-2sources} illustrates a request made to the server on a
control channel and a reply being received on a broadband data channel.
This sort of architecture could be used for controlling the delivery of
``pay--per--view'' television, where the RPC mechanism is used by an
application to make a payment over a control network and the data is
delivered on differently constructed channels over a broadband network
possibly to different hardware.

Essentially the differences between the two architectures, with regard to
the activation of the channel objects, are:
\begin{itemize}
\item synchronizing with the call, and
\item the opacity of the call's contents
\end{itemize}

A stream--oriented architecture need not transmit data to the server in
a contiguous block that represents the marshalled bytes of a message. A
call--oriented architecture would have each channel object invoked with
each call sent.

A stream--oriented architecture only has access to the marshalled bytes
that represent a message. A call--oriented structure sees the method that
is invoked and the parameters for it.

\begin{figure}[htbp]
  \begin{center}
  \includegraphics[keepaspectratio=1, totalheight=4 in]{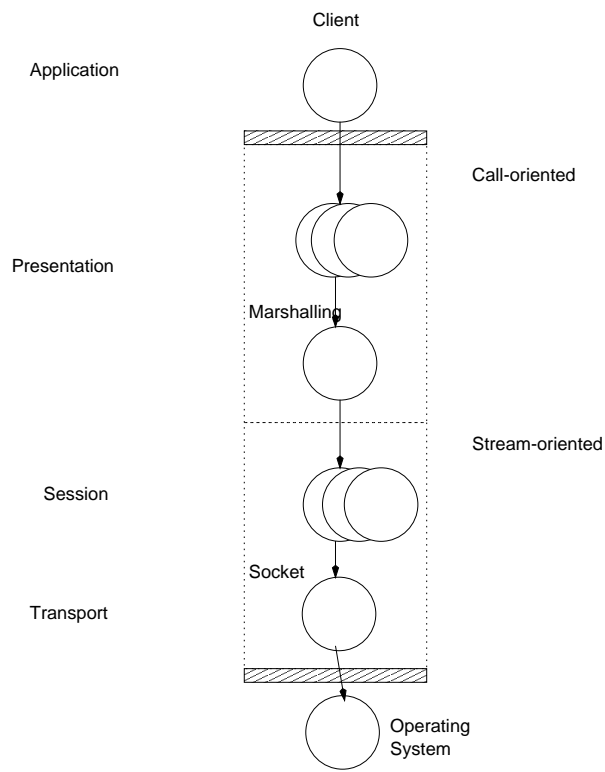}
  \caption{The Marshalling Object: the boundary between call-- and
    stream--oriented architectures}
  \label{fig:odp-call-stream}
  \end{center}
\end{figure}

The most flexible architecture is the call--oriented one. But for
implementing the transport objects, a stream--oriented architecture should
be preferred. This means that the two architectures fall above and below
the marshalling channel object, see figure \ref{fig:odp-call-stream}.

This figure attempts to place the channel objects in an Open Systems
Interconnection model, \cite{OSI:intro}. Objects in the call--oriented
architecture provide presentation layer services and, after the marshalling 
object, which reduces a call to a sequence of bytes, the session objects
and finally the network transport object, a socket driver, can operate upon 
the data as a stream. The session layer objects would also perform
stream--oriented encryption, but a presentation layer object would
negotiate keys.

Channel objects that are call--oriented will be called
\emph{call--handlers} and those that are stream--oriented will be called
\emph{stream--handlers}.

The marshalling object is a call--handler and, if need be, it could invoked 
a number of time to render the data as a sequence of bytes. This would be
useful for data security, since it may be necessary to encrypt the data and 
make it unintelligible to other call--handlers.

\section{Stream--Oriented Architecture: Design}
\label{sec:stream-oriented:design}

\textit{Sun} with \Java and the socket factory technique have implemented
what is described later as a simplex system, \myRef{sec:simplex}: each
stream has two channel objects, one for sending---the output stream---and
one for receiving---the input stream. There are two kinds of pairs: the
client's, or, more precisely, the initiator's, pair and the server's, or
acceptor's, pair.

Although there is no explicit code to synchronize the two streams so that
only one may be used at a time, it is not expected that an application can
simultaneously send and receive. This limitation could be removed by a
suitably designed channel object which could specify a different return
address.

\section{Call--Oriented Architecture: Design}

As pointed out above, a message passed between a sender and receiver would
negotiate four different channels:

\begin{enumerate}
\item Request
\item Indication
\item Response
\item Confirmation
\end{enumerate}

but if the call is a cast of some kind, it need only have two:
\begin{enumerate}
\item Request
\item Indication
\end{enumerate}
  
And some parts may not do anything, for example a request handler could log
all calls made by the client, but the server need not record all the calls
that are made upon it.

When an application programmer makes use of a remote procedure call the
infrastructure sends the message to the server and blocks the thread
pending the arrival of the confirmation. One could also implement the
channel objects in this way. This leads to two different architectures:

\begin{itemize}
\item Either: one channel object for each of request, indication, response
  and confirmation \- \emph{Simplex}.
\item Or: two channel objects: one that requests and blocks pending the
  arrival of a confirmation; one that receives indications, invokes the
  service (and blocks waiting for it) and then sends the response \-
  \emph{Duplex}.
\end{itemize}

The latter will be discussed first.

\subsection{Duplex: One Pair of Objects}

One channel object for the initiator performs request and confirmation,
another channel object performs indication and response for the acceptor.
This means that every channel object has a bi--directional data--flow with
the channel object below it, rather like the application object has with
the distinct channels in figure \ref{fig:odp-4channels}. This has the
attraction that should the initiator's channel objects for the request and
the confirmation need to share state, then this is achieved implicitly
because they are the same object.

This has a problem with broadcasts, (or one--casts). The initiator's
channel object (request and confirmation) needs to determine if it is to
block or not and that would require this information be made available to
the channel object, perhaps best supplied as a parameter to the method
invocation. This information is required in \textit{CORBA}, the Interface
Definition Language allows methods on interfaces to be marked as \mySrc{one
  way}, but is not part of the \Java specification; although it could be
inferred if the definition of the remote method returns no result, in which
case, the method can be invoked as a one--cast.  However, if exceptions are 
returned by the remote server then the method is a call: the exception,
whether raised or not, is a response.

\subsection{Simplex: Two Pairs of Objects}
\label{sec:simplex}

If a channel object is implemented for each stage of the communication,
then there are, at most, two pairs of objects: a request and confirmation
pair at the initiator; an indication and response pair at the acceptor.
Figure \ref{fig:odp-4channels} illustrates the engineering of this.

Because the objects are distinct they will need to bind with one another
should they need to share state: an example of the difficulties this might
lead to can be seen when communicating and attempting to correct errors.

The infrastructure would block the application programmer's thread of
control pending the arrival of the confirmation message from the server
containing the results of the call.  Ordinarily, the confirmation channel
objects will unravel the message returned by the server, instantiate the
result for the application programmer's thread and unblock it.

If there is an error in any of the channel objects, then it should be
possible for the channel objects to attempt to clear the error and resend
the message if need be. The problem then is that the channel objects in the
request channel need to be synchronized with the error results received in
the confirmation channel.  If there is a failure in one of the channel
objects in the indication channel, then they must be propagated via the
response and confirmation channels. This is a difficult issue and is
discussed later at greater length, \myRef{sec:exceptions}.

The important difference between the duplex and simplex methods is the
relationship to the state of the call. With the duplex model the request
channel retains the state of the call awaiting an acknowledgement from the
confirmation channel. What makes the duplex model retain state is that the
channel objects invoke one another and form a stack of calls which can be
associated with a thread.

This can be emulated in the engineering of a simplex model without
requiring the use of the stack, by passing a call identifier. This would
only be of use if the channel objects were also to retain their internal
state when they release control.

\paragraph{Example: A Secured Message Delivery Service}

As an example, an encryption and decryption service would require:
\begin{enumerate}
\item Request \- encrypt
\item Indication \- decrypt
\item Response \- encrypt
\item Confirmation \- decrypt
\end{enumerate}

There are only two functions---encrypt and decrypt---but performed at four
locations. It should be possible to provide just one pair of
implementations---an \mySrc{Encryptor} and a \mySrc{Decryptor}---located
differently.
\begin{enumerate}
\item \mySrc{Encryptor}
  \begin{itemize}
  \item Request channel object \- initiator
  \item Response channel object \- acceptor
  \end{itemize}
\item \mySrc{Decryptor}
  \begin{itemize}
  \item Indication channel object \- acceptor
  \item Confirmation channel object \- initiator
  \end{itemize}
\end{enumerate}

\mySrc{Encryptor} and \mySrc{Decryptor} would both be implemented as
stream--handler channel objects.

Encryption is subject to replays of old messages unless the messages are
time--stamped and sequenced. Usually, time--stamping and sequencing are
implemented as part of the encryption and decryption objects, but with
channel objects they can be provided separately. Time--stamping has two
functions:

\begin{enumerate}
\item Request \- timestamp issue
\item Indication \- timestamp check
\item Response \- timestamp issue
\item Confirmation \- timestamp check
\end{enumerate}

Two functions: \mySrc{StampIssuer}, \mySrc{StampChecker}; four locations:

\begin{enumerate}
\item \mySrc{StampIssuer}
  \begin{itemize}
  \item Request channel object \- initiator
  \item Response channel object \- acceptor
  \end{itemize}
\item \mySrc{StampChecker}
  \begin{itemize}
  \item Indication channel object \- acceptor
  \item Confirmation channel object \- initiator
  \end{itemize}
\end{enumerate}

Similarly for a sequence number generator and checker.

\mySrc{StampIssuer} and \mySrc{StampChecker} would both be implemented as
call--handler channel objects.

Because it is difficult to synchronize clocks in a distributed networks,
some secure message delivery systems allow some skew on the clocks and use
checksums to detect replayed messages. Only the acceptor's indication
channel and the initiator's confirmation need deploy a
\mySrc{ReplayDetector} object. It would be implemented as a call--handler
generating a checksum from the marshalled data incoming as an indication
or as a confirmation.

\subsection{Simplex or Duplex}

There is a greater similarity in the simplex architecture to the
engineering underlying the messaging system than in the duplex
architecture, but the duplex architecture has some attractive
state--retention properties which should be emulated in a simplex
architectureq. The rest of this discussion will concern itself with a
simplex architecture that attempts to retain state across all four phases
of a call.

The other great attraction of the simplex architecture is that if the
channel objects reside in different channels, then it is easier to relocate
the channel. This would be especially useful when a call uses two different
media as the example system in figure \ref{fig:odp-2sources} illustrated.


\section{Service Invocation Semantics}

From what has been said above, the \Java RPC mechanism has a
stream--oriented architecture and what follows is a proposed call--oriented
architecture for it. It should serve as an example of how channel objects
could be deployed in other \textit{CORBA} RPC systems.

The distributed processing infrastructure will have to determine the order
in which the channel objects will be invoked. There are now two ways in
which the methods of the channel objects can be invoked and how they should
bind with one another.

\begin{itemize}
\item Either: have the request object invoke a method on the indication
  object and block waiting for the response: \emph{Peer--to--Peer}
  invocation.
\item Or: have the request object perform its work and return a modified
  call object and return immediately: \emph{Service} invocation.
\end{itemize}

The peer--to--peer option requires the duplex architecture which has
been dismissed, so only the service method of invocation is left.
Peer--to--peer is very attractive, it would allow channel objects to
be client--server pairs and thus be able to use the stub--compiler.
Unfortunately, it would demand too much memory to stack all of the calls
required to navigate complicated protocol stacks.

The approach taken by \textit{Sun} in their own implementation of
\mySrc{java.rmi.server.RemoteRef}, shipped as
\mySrc{sun.rmi.server.UnicastRef} as part of the \Java Runtime Environment,
is suitable for the simplex architecture proposed. Referring to the
the code fragment given in the appendix \myRef{sec:stub}, the key method is
\mySrc{java.rmi.server.RemoteRef.invoke()}. It is implemented along the
lines given in the next code fragment: note the two comments
indicating when the two types of channel objects should be invoked.

\begin{verbatim}
package sun.rmi.server;

public class UnicastRef implements RemoteRef {

  public Object invoke(Remote r, reflect.Method m,
                       Object[] p, long l) throws Exception
    {
      // *Request channel objects should be invoked now*

      // Establish a connection using information in Remote r

      // Get the streams associated with the connection
      // Marshall data onto the stream
      // Execute the call
       
      // *Confirmation channel objects should be invoked now*

      // Unmarshall
      // Release the connection
      // Return the result
    }
}

\end{verbatim}

\subsection{Wrappers: Request and Response Channels}

\begin{figure}[htbp]
  \begin{center}
  \includegraphics{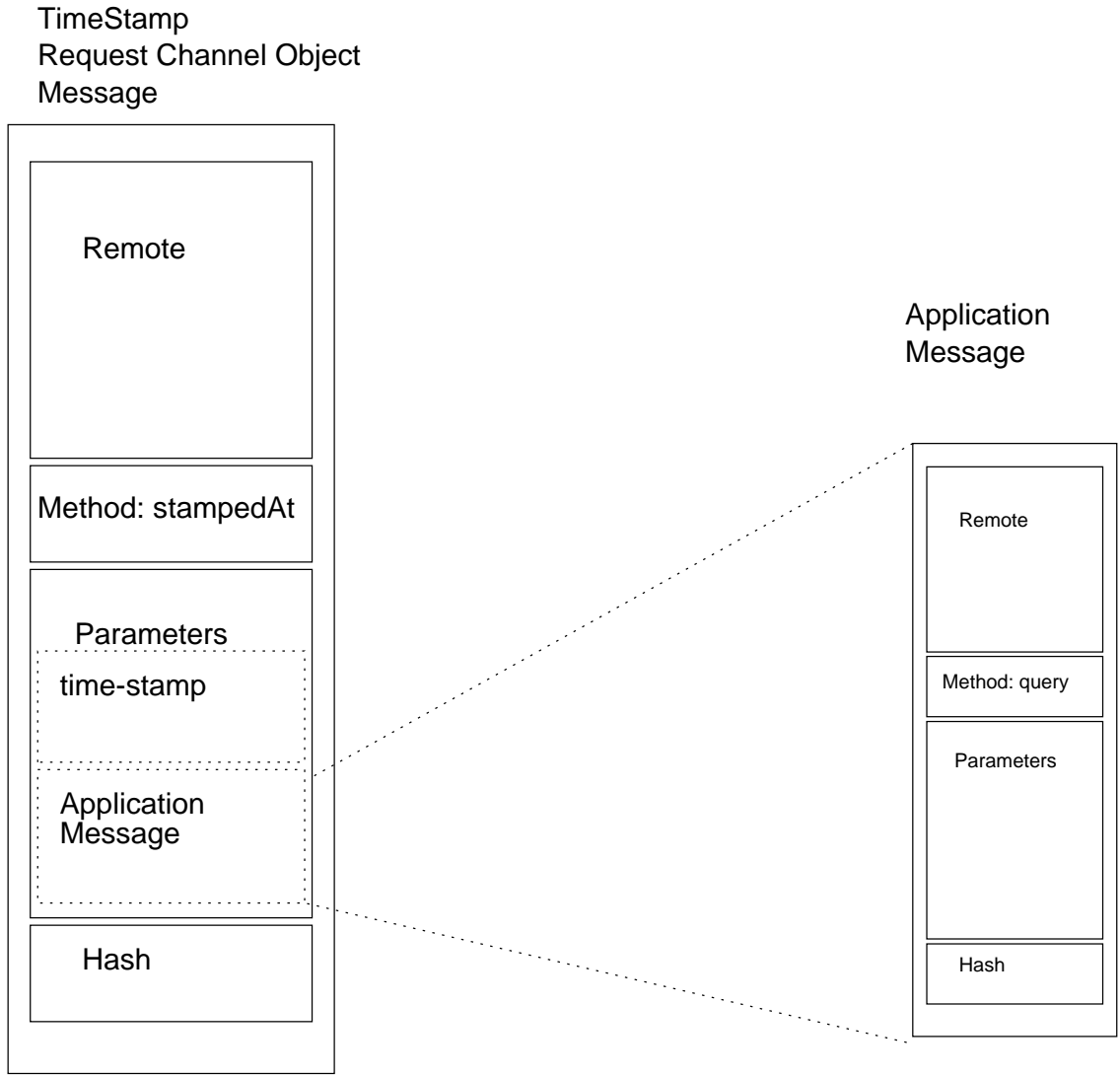}
  \caption{A message containing another message}
  \label{fig:odp-channel-message-1}
  \end{center}
\end{figure}

The channel objects would be invoked serially and would be passed the same
parameters as \mySrc{invoke()}, collectively call these a \mySrc{Message}
object. The request channel objects would return a \mySrc{Message} object,
but these would usually have the original message as one of its parameters.
This is a simple encapsulation procedure and is illustrated in figure
\ref{fig:odp-channel-message-1}, where a time--stamping channel object has
been passed the application object's message. The application object wants
to invoke method \mySrc{query()}, the channel object passes this message as
a parameter of its own message. That message invokes method
\mySrc{stampedAt()}. The two message objects might duplicate target
and return addresses, it should be possible to remove this redundancy, but
it should still be possible to specify a different target address and a
different return address if need be.

\subsection{Unwrappers: Indication and Confirmation Channels}

The objects located in the indication and confirmation channels would need
a skeleton rather like that described above, \myRef{sec:inflexible}, either
a dispatcher or use a reflective invocation on the channel object's service
implementation. After the service has completed processing, it would return
its results to the skeleton which would then return the message object to
the infrastructure.

\subsection{Counterparts and Associates}

Some channel objects will have counterparts in the remote channel, for
example, a time--stamping request channel object should have a time--stamp
checking indication channel object, but some may not, for example a request
logging channel object placed in a server's indication channel. If a
channel object sends then its counterpart receives and \textit{vice--versa}.

A channel object can have an associate in a local channel. A request 
channel object could have an associate in the confirmation channel. If a
channel objects sends then its associate receives and \textit{vice--versa}.

\begin{figure}[htbp]
  \begin{center}
  \includegraphics[angle=-90, keepaspectratio=1, totalheight=6 in]{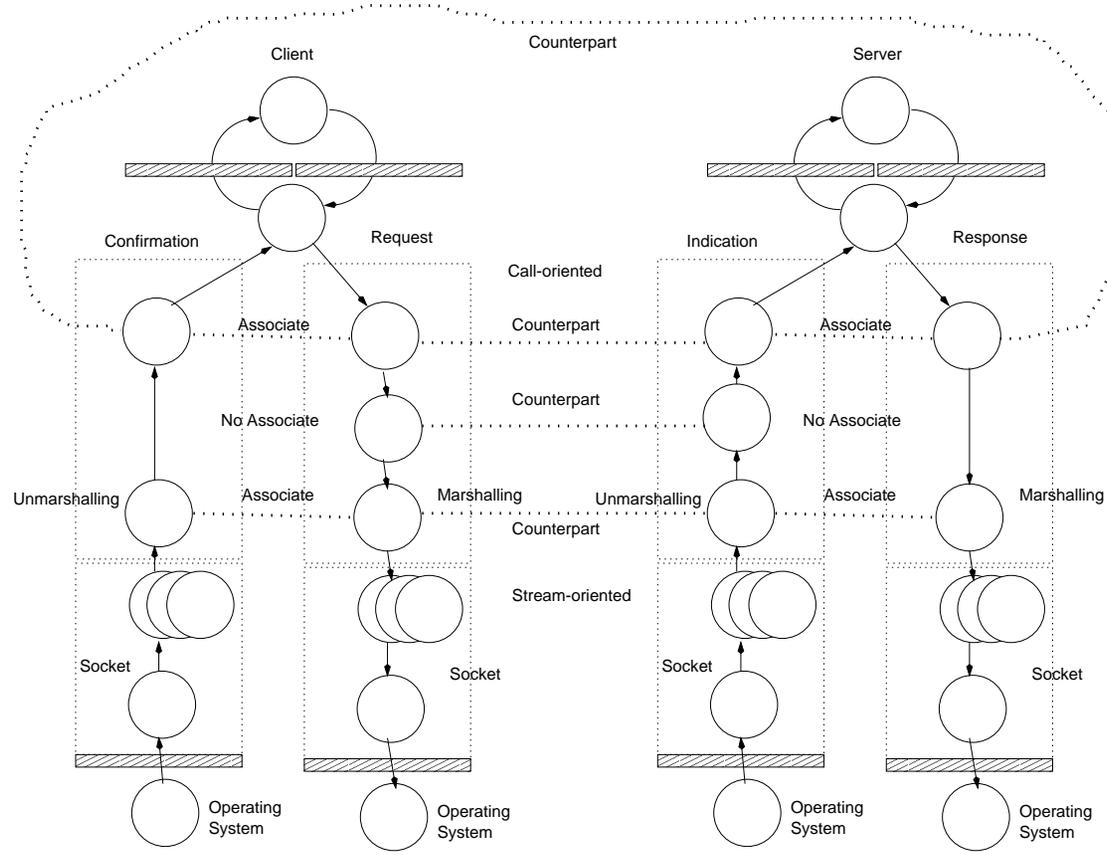}
  \caption{Counterpart and Associate Channel objects}
  \label{fig:odp-counterparts}
  \end{center}
\end{figure}

Figure \ref{fig:odp-counterparts} should help to clarify this.

\begin{enumerate}
\item Marshalling and Unmarshalling
  
  Looking at the marshalling and unmarshalling objects: the marshalling
  object in the client's request channel has a counterpart in the server's
  indication channel and an associate in the client's confirmation channel.

\item Timing

  The top layer of channel objects are used for timing: time--stamping and
  time--checking, they comprise a full complement, where each channel
  object has an associate and a complement.

\item Usage

  The second layer of channel objects is used to record usage statistics
  and only logs requests and responses, so they have no associates but a
  counterpart.
  
\end{enumerate}

Clearly, this might prove to be cumbersome, it might be simpler to insist
that all channel objects have a counterpart and an associate and have a
default implementation which just copies the message over.

\subsection{Exception Handling}
\label{sec:exceptions}

Any of the channel objects can raise an exception, but part of the function
of channel objects is to attempt to clear exceptions, for example:
\begin{itemize}
\item relocation objects would obtain new addresses for servers,
\item key management objects would obtain new keys in the event of expiry.
\item authorization managers could obtain new privileges.
\end{itemize}

The channel objects have to retain some state that would allow them to act
upon exceptions. This would require that they place state in a common
object that associated channel objects can access. The state would need to
be stored with an identifier unique to the call.

\begin{enumerate}
  
\item Exception Handling in One Channel

  \begin{enumerate}
  \item Clear
  
    When one channel object raises an exception, the infrastructure signals
    the other channel objects in the same channel, in the reverse order in
    which they were invoked, asking them to attempt to clear the exception.
  
  \item Uncleared: Undo
  
    If the exception cannot be cleared then the channel objects should be
    signalled to undo their previous actions.
  
  \item Cleared: Undo and Redo
  
    If the exception can be cleared then the channel objects may need to
    undo their previous actions and be allowed to redo them.

  \end{enumerate}
  
  Redo is a distinct operation, because it may allow the implementation to
  be optimized for error recovery.

  Co--ordinating this is a little difficult. There are two sequences:
  \begin{enumerate}
  \item Attempt to clear then undo and then redo
    
    Each object in turn in ascending order (\ie reverse order to
    invocation) attempts to clear the exception, if any one succeeds then
    the undo operation is invoked in ascending order to the top of the
    channel. Then the redo operation is invoked in descending order.
    
  \item Attempt to clear and undo and then redo
    
    Each object in turn in ascending order (\ie reverse order to
    invocation) attempts to clear the exception and performs an undo. If
    any one succeeds then the undo is invoked on the remaining objects in
    ascending order and then the redo operation is invoked in descending
    order.
    
  \end{enumerate}
  
  The former might prove more efficient if errors are expected to be
  cleared, the latter if not. The idea is communicated in figure
  \ref{fig:odp-exceptions}, but the details of invocation are not.

  \begin{figure}[htbp]
    \begin{center}
      \includegraphics{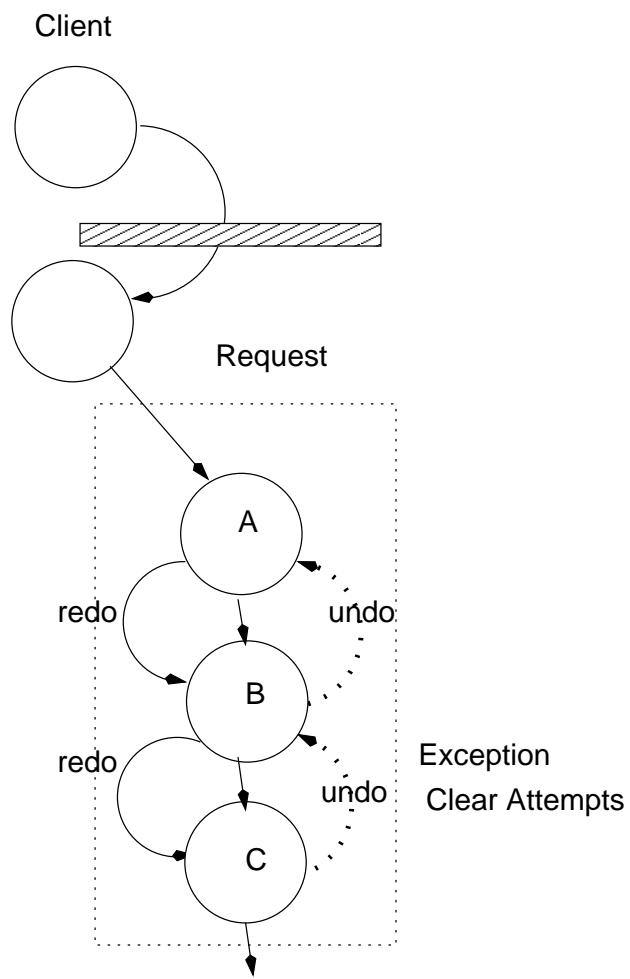}
      \caption{Exception handling in one channel}
      \label{fig:odp-exceptions}
    \end{center}
  \end{figure}
  
  If the latter scheme (Attempt to Clear and Undo Simultaneously) is used:
  channel object $C$ raises an exception, it undoes its action, passes back
  the original message it received to object $B$ which also undoes its
  action and returns the original message it received to $A$. $A$ clears
  the exception and redoes its action and passes the message on to $B$
  which also redoes its action and thence to $C$.

\item Exception Handling Across Channels
  
  If a receiving counterpart raises an exception, it should be signalled to
  the sender: this would be achieved by the receiving counterpart sending a
  message to its sending associate to raise an exception with its receiving
  counterpart. \Java already has a proven mechanism for this, objects of
  class \mySrc{Exception} can be contained in objects of class
  \mySrc{RemoteException}.

  \begin{enumerate}
    
  \item Exception raised in the Indication Channel
      
    As an example, see figure \ref{fig:odp-exceptions-1}, a request channel
    object sends a message which raises an exception in the indication
    channel, the channel object in the indication channel raises an alert
    with its associate in the response channel, which sends an exception to
    its counterpart in the confirmation channel.
    
    \begin{figure}[htbp]
      \begin{center}
        \includegraphics[angle=-90, keepaspectratio=1, totalheight=6 in]{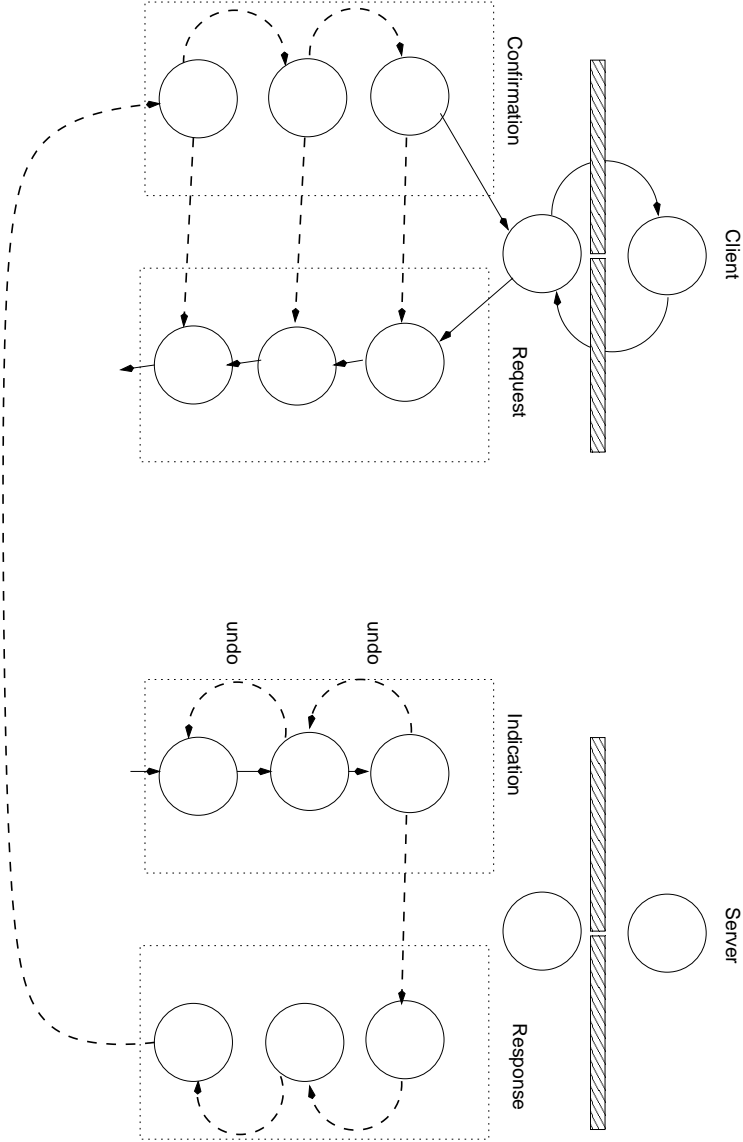}
        \caption{Exception handling across channels}
        \label{fig:odp-exceptions-1}
      \end{center}
    \end{figure}

    Points to note in figure \ref{fig:odp-exceptions-1} are:
    \begin{enumerate}
    \item The indication channel is cleared down
    \item The server receives no message
    \item The response channel propagates the exception
    \item The objects in the confirmation channel can signal their
      associates in the request channel.
    \end{enumerate}

    It might be possible for the request channel objects to invoke the same 
    procedure as portrayed in figure \ref{fig:odp-exceptions}, clear the
    exception and re--send the message, this would require that they have
    access to the message as issued by the client.

  \item Exception raised in the Confirmation Channel
      
    If a response channel object sends a message which raises an exception
    in the confirmation channel, then the confirmation channel object would
    signal its associate in the request channel, which may, if it has
    retained state, \ie the message it sent, be able to clear the exception
    and re--send without intervention by the client application object.
      
  \end{enumerate}
  
\end{enumerate}

\subsection{Interfaces and Classes}

\subsubsection{A \mySrc{Message} Class}

A simple message class is needed to contain the parameters passed to
the \mySrc{invoke()} method and to all the channel objects.

\begin{verbatim}
package java.lang.rmi.channel;

import java.rmi.*;
import java.lang.reflect.*;

public class Message {
  Remote remote;
  Method method;
  Object[] parms;
  long hash;

  public Message(Remote r, Method m, Object[] p, long h) {
    remote = r;
    method = m;
    parms = p;
    hash = h;
  }
}
\end{verbatim}

\subsubsection{Channel Objects}

\paragraph{Basic Interface}

Channel objects then all have the same interface and it is their location
which determines their function, \ie whether they wrap or unwrap.

\begin{verbatim}
package java.lang.rmi.channel;

public interface Handler {
  public Message clear(Message m, Exception e) throws Exception;
  public Message todo(Message m) throws Exception;
  public Message undo(Message m, Exception e) throws ClearedException;
  public Message redo(Message m) throws Exception;
}
\end{verbatim}

Both methods of clearing exceptions are covered in this interface because
there is a separate \mySrc{clear()} method.

\paragraph{Exceptions}

The \mySrc{undo()} method throws an exception to indicate it has cleared
the exception it was passed.

\begin{verbatim}
package java.lang.rmi.channel;

public class ClearedException extends Exception {

    public ClearedException(String s) {
      super(s);
    }

    public ClearedException(String s, Exception ex) {
      super(s, ex);
    }

}
\end{verbatim}

Other exceptions which might make processing more decisive:

\begin{enumerate}
\item Unclearable 
  
  It might also prove useful to have \mySrc{clear()} raise an exception
  that the exception cannot be cleared and this would allow the
  infrastructure to request a re--send decision from the application
  object.

\item Rebind
  
  It might also prove useful if the channel objects can demand a rebind and 
  force the destruction of their channel. This would be useful for a
  relocation channel object.

\end{enumerate}

\paragraph{Handler Specification}

The next is an abstract class that provides a container to hold the channel
objects which would be loaded by the infrastructue. It might be wise, for
security reasons, to implement the class more fully and make the
\mySrc{getHandler()} method final. The class also provides a means of
obtaining associates and counterparts.

\begin{verbatim}
package java.lang.rmi.channel;

import java.rmi.*;

public abstract class Handlers 
{
  public static final int REQUEST = 1;
  public static final int INDICATION = 2;
  public static final int RESPONSE = 3;
  public static final int CONFIRMATION = 4;

  public Handlers() {
    ;
  }

  Handler getHandler(int identity) {
    return null;
  }

  Remote getCounterPart(int identity) {
    return null;
  }

  Object getAssociate(int identity) {
    return null;
  }
}
\end{verbatim}

\paragraph{Channels}

Channels themselves would be simple containers probably implemented with a
vector object which can be easily iterated in the \mySrc{invoke()}
operation.

\begin{verbatim}
package java.lang.rmi.channel;

import java.rmi.*;

public abstract class Channel 
{
  private int identity = -1;
  private Handlers handlers;

  public Channel(int id, Channels h) {
    identity = id;
    handlers = h;
  }
}
\end{verbatim}

\subsubsection{Channel Objects: Binding and Invoking Methods}

\paragraph{Binding}

Channel objects would need to bind with one another, so that they can
exchange parameters. The channel object methods are defined on one
interface, they can therefore suppport other methods on another
service interface.

\paragraph{Invocation}

When a channel object is passed a message object, it will need to invoke a
method on its counterpart. The easiest way for it to do this is to use the
same stub and skeleton mechanism for dispatching a call as remote procedure 
calls use. As can be seen in figure \ref{fig:odp-channel-message-1}, where
one channel object adds enough parameters to the message so that when it is 
received by its counterpart, it can use a local \mySrc{invoke()} method to
pass the parameters and obtain the resulting message object from a local
service interface.

If channel objects have to invoke methods upon one another they can either
send them with the message or they can go ``out--of--band'' and communicate 
them directly if they support the \mySrc{java.rmi.Remote} interface using
their own channel.

\section{Summary}

\subsection{Previous Work}

\Java certainly has the functionality to implement channel objects. The
author has already implemented a similar scheme (using a duplex,
peer--to--peer architecture) for a variant of the \textit{ANSAware},
\cite{web:ansa}, known as \textit{DAIS}, the distributed processing
platform produced by \textit{ICL}, \cite{web:icl}, which was \textit{CORBA}
compliant. That implementation, written wholly in \textit{C},
\cite{Kernighan:CPL78}, was able to support a \textit{Kerberos},
\cite{sec:kerberos}, authentication and confidentiality service invisibly
to the application programmer, the channels were constructed according to a
template specified by an environment variable.

The author has developed a prototype application which can support the
Transport Level Security protocol. (A fairly rigorous treatment of both it
and the \emph{Kerberos} protocol is given in \cite{tls:eaves}.) A simple
channel object is put in place that negotiates the keys, whilst a custom
socket is used to perform the encryption.

\textit{ANSAWare} evolved to a product known as \textit{Reflective Java},
\cite{sft:reflect:java}, which supported channel objects, but was designed
to provide support to application objects---presentation of data and so
forth---and not to provide system services.

\subsection{Difficulties of Binding}

The principal difficulty faced in these prototypes is that there is no
simple or well--defined mechanism for expressing how two parties should
bind with one another. \textit{ANSAware} had a well--developed binding
model, \cite{sft:ansaware}, but its final implementation had a limited
number of quality of service parameters.

\Java now has a very well--developed security architecture,
\cite{sec:java:security}, but currently seems to have no means of
specifying the degree of security \emph{required}, it seems to be
principally oriented towards setting permissions. (Of course, a permission
one does not have is a requirement.)  The \mySrc{SecurityManager} would
appear to be the best--placed component of the \Java security architecture
to negotiate and configure channels. It will be difficult for any
distributed processing platform to achieve any degreee of acceptance in an
enterprise--wide data processing environment if it is not possible to
implement many of the logging services that have long been available in
mainframe systems. The most pertinent example of which is \textit{CICS}
\cite{Wipfler87}, the Customer Information Control System, which was
originally a messaging system, but was extended to become a transaction
monitor that could manage database enquiries.

Hopefully, with channel objects in place, remote procedure calls could do
the same across the World--Wide Web.


\appendix

\section{Stubs for the RMI of \Java}
\label{cha:stubs}

The Remote Method Invocation package of \Java needs to use a
stub--compiler to generate stubs. It no longer needs to generate skeletons
for servers.

There is, of course, no need to distribute the stubs, the client can
collect them from the server.

\subsection{Interface and Implementation}
\label{sec:interface}

\paragraph{Interface}

This very simply returns a string, given a string.

\begin{verbatim}
package rmi.demo;

public interface Answerer extends java.rmi.Remote {
  String answer(String mesg) throws java.rmi.RemoteException;
}
\end{verbatim}

\paragraph{Implementation}

This is the implementation of the interface. Most of the code involves
posting the server's reference to the Naming service.

\begin{verbatim}
package rmi.demo;

import java.rmi.*;
import java.rmi.server.UnicastRemoteObject;

public class AnswererImpl
  extends UnicastRemoteObject
  implements Answerer
{
  private String name;

  public AnswererImpl(String s) throws RemoteException {
    super();
    name = s;
  }

  // Begin
  // Remotely accessible methods
  // Parameters have to be implementations of Serializable
  
  public String answer(String message) throws RemoteException {
    System.out.println("Received:" + message);
    return new String("You said:" + message);
  }

  // End

  public static void main(String args[])
  {
    // Create and install a security manager
    System.setSecurityManager(new RMISecurityManager());
    // Without it no new classes can be loaded

    try {
      System.out.println("Construct");
      AnswererImpl obj = new AnswererImpl("AnswererServer");
      System.out.println("Bind");
      Naming.rebind("AnswererServer", obj);
      System.out.println("Bound");
    } catch (Exception e) {
      System.out.println("AnswererImpl err: " + e.getMessage());
      e.printStackTrace();
    }
  }
}

\end{verbatim}

\subsection{Stub}
\label{sec:stub}

This is the version 1.2 stub for the \mySrc{rmi.demo.Answerer} service
interface.

\begin{enumerate}
\item Reflective Language Features

  The most important feature of these stubs is that they make use of the
  capabilities provided by \mySrc{java.lang.reflect}. When the class is
  loaded by the client, and it is designed to be loaded from a remote
  location, the initialization code instantiates the
  \mySrc{java.lang.reflect.method} attribute for the stub
  \mySrc{\$ method\_answer\_0} using the \mySrc{getMethod()} method of
  \mySrc{java.lang.Class}.

\item Instantiating objects that implement \mySrc{rmi.demo.Answerer}
  
  Because the client has to use the naming service to locate an object that
  implements the \mySrc{rmi.demo.Answerer}, the naming service loads the
  stub class and calls the constructor in it
  \mySrc{rmi.demo.AnswererImpl\_Stub()}. The client casts the object the naming
  service returns to be \mySrc{rmi.demo.Answerer}.

\item Invoking methods
  
  The stub compiler generates proxy implementations of the methods that are
  declared in \mySrc{rmi.demo.Answerer}. They use the reflective method
  variable \mySrc{\$method\_answer\_0} and the
  \mySrc{java.rmi.server.RemoteRef.invoke()} method.
  
\end{enumerate}

\begin{verbatim}
// Stub class generated by rmic, do not edit.
// Contents subject to change without notice.

package rmi.demo;

public final class AnswererImpl_Stub
  extends java.rmi.server.RemoteStub
  implements rmi.demo.Answerer, java.rmi.Remote
{
  private static final long serialVersionUID = 2;
    
  private static java.lang.reflect.Method $method_answer_0;
    
  static {
    try {
      $method_answer_0 =
        rmi.demo.Answerer.class.getMethod("answer",
                                          new java.lang.Class[]
                                          {java.lang.String.class});  
    } catch (java.lang.NoSuchMethodException e) {
      throw new
        java.lang.NoSuchMethodError("stub class initialization failed");
    }
  }
    
  // constructors
  public AnswererImpl_Stub(java.rmi.server.RemoteRef ref) {
    super(ref);
  }
    
  // methods from remote interfaces
    
  // implementation of answer(String)
  public java.lang.String answer(java.lang.String $param_String_1)
    throws java.rmi.RemoteException
    {
      try {
        Object $result =
          ref.invoke(this, $method_answer_0,
                     new java.lang.Object[] {$param_String_1},
                     -8351992698817289230L);
        return ((java.lang.String) $result);
      } catch (java.lang.RuntimeException e) {
        throw e;
      } catch (java.rmi.RemoteException e) {
        throw e;
      } catch (java.lang.Exception e) {
        throw new
          java.rmi.UnexpectedException("undeclared checked exception",
                                       e);
      }
    }
}

\end{verbatim} 


\section{Funding and Author Details}

Research was funded by the Engineering and Physical Sciences Research
Council of the United Kingdom. Thanks to Malcolm Clarke, Russell--Wynn
Jones and Robert Thurlby.

\begin{verse}
  Walter Eaves \\
  Department of Electrical Engineering, \\
  Brunel University \\
  Uxbridge, \\
  Middlesex UB8 3PH, \\
  United Kingdom
\end{verse}

\begin{verse}
  \url{Walter.Eaves@bigfoot.com} \\
  \url{Walter.Eaves@brunel.ac.uk}
\end{verse}

\begin{verse}
  \url{http://www.bigfoot.com/~Walter.Eaves} \\
  \url{http://www.brunel.ac.uk/~eepgwde}
\end{verse}

\sloppy

\end{document}